\documentclass{article}
\usepackage[accepted]{vietnam} 
\usepackage{natbib}
\usepackage{graphicx}      
\begin{document}
\twocolumn[
\title{Multi-line Imaging of the Starburst Galaxy NGC 1808 with ALMA}

\author{Dragan SALAK, Yuto TOMIYASU, Naomasa NAKAI, and Yusuke MIYAMOTO}{d.salak@kwansei.ac.jp}

\address{Department of Physics, Kwansei Gakuin University, 2-1 Gakuen, Sanda, 669-1337 Hyogo, Japan}

\vskip 0.5cm 
]

\begin{abstract}
Revealing the properties of molecular gas in starburst galaxies is key to understand the process of star formation across cosmic history. In this paper, we present our recent high-resolution (\(\sim1''\) or 50 pc) ALMA observations of the nearby, barred starburst galaxy NGC 1808, and highlight some of the main findings: (1) the discovery of a double peak revealed in the distributions of CO (3-2) and CS (2-1) in the circumnuclear disk indicating the presence of a molecular gas torus with a radius of \(r\sim30\) pc; inside the torus, we found a compact source of dust continuum at 0.87 mm and CO, coincident with the location of the low-luminosity AGN revealed by X-ray observations; (2) the excitation of CO gas, traced by the line intensity ratio of CO (3-2) to CO (1-0) is elevated to \(\sim1\) in the star-forming disk; the ratio is between 0.3 and 1 in the 500 pc ring; (3) the ratio of HCN (1-0) to HCO\(^{+}\) (1-0) in the central 1 kpc exhibits a radial gradient: while the ratio decreases from \(\sim1.5\) in the center to \(\sim0.9\) in the 500 pc disk, it exhibits another peak of \(\sim1.2\) in a ring-like structure at a galactocentric radius of \(r\sim300\) pc. The detection of SiO (2-1) in the circumnuclear disk suggests the presence of shocks, likely generated by intense nuclear star formation (supernova explosions and winds from massive stars), as well as inflows and outflows of turbulent gas in the central 100 pc.

\end{abstract}

\section{Introduction}

Star formation (SF) plays a key role in the evolution of galaxies. Observations of distant galaxies reveal that the cosmic star formation rate (SFR) density reached its maximum at redshift \(z\sim2\), when the Universe was only a few billion years old \cite{MD14}. There is evidence that the conditions of SF and interstellar medium (ISM) in this dynamic epoch resembled the conditions found in starburst galaxies, where SFR can be so large (\(10\sim10^3~M_\odot~\mathrm{yr}^{-1}\)) that the depletion time (needed to turn all molecular gas into stars) is a tiny fraction of the host galaxy's lifetime \cite{Tac13}. Compared to the clouds in typical galactic disks, the molecular gas in starburst environments exhibits extreme values of surface density, velocity dispersion, and pressure. In order to understand the SF in galactic central regions and especially the formation of massive stars and star clusters, it is essential to understand the properties of molecular clouds in starburst galaxies.

Before Atacama Large Millimeter/submillimeter Array (ALMA), detailed observational studies of starburst galaxies at high resolution (\(<100\) pc) and sensitivity were limited to few nearby objects. The situation has now changed, and this work represents an example of observations of a starburst galaxy using ALMA in its early operations (cycle 2). The target is NGC 1808, a nearby (11 Mpc), barred starburst galaxy with peculiar polar dust lanes and gas outflow emerging from the nucleus (Fig. 1a) \cite{Phi93,Sal16}. In this presentation, we outline the observations and focus on: (1) the discovery of a molecular gas torus in the galactic center (central 100 pc) traced with \(^{12}\)CO (\(J=3-2\)) and CS (2-1), (2) the intensity ratio of the CO (3-2) and CO (1-0) lines at 100-pc scale, (3) the distribution of dense gas tracers HCN (1-0) and HCO\(^{+}\) (1-0), whose intensity ratio exhibits a radial gradient in the central 500 pc disk, and (4) the first detection of SiO (2-1) in NGC 1808. Details about the observations and data reduction, as well as a more comprehensive analysis will be presented in a forthcoming paper (Salak et al. in prep).

\section{Results and Discussion}

\subsection{Molecular gas torus and core in the central 100 pc}

The central 1 kiloparsec of NGC 1808 observed in CO (3-2) at 345.8 GHz by ALMA is shown in Fig. 1b. CO was detected throughout the region at unprecedented angular resolution of \(0.6\arcsec\) (\(\sim30\) pc), with a sensitivity of 8 mJy beam\(^{-1}\) in a channel of 5 km s\(^{-1}\), revealing the distribution of molecular clouds in the circumnuclear disk (CND; central 100 pc), 500 pc pseudoring, and a SF disk including a spiral pattern between the CND and the 500 pc ring (Fig. 1b,c). As expected, the structure resembles the distribution of CO (1-0) reported in our previous work \cite{Sal16}. The new image shows that molecular gas in the central 400 pc disk is organized in a multi-arm pattern and giant molecular clouds (denoted by SF GMCs in Fig 1c). Figure 1d shows the distribution of thermal (0.87 mm continuum) emission that originates in dust grains. Detected mostly from compact sources (SF GMCs), the distribution correlates very well with CO; these dusty GMCs are the nurseries of newly-formed massive star clusters revealed in infrared observations (e.g., \citealt{Bus16}). The major difference between the continuum and CO is observed in the CND (central 100 pc; Fig. 1e). In this region, we reveal for the first time that molecular gas exhibits a double peak, whereas the dust continuum is strongest between the CO peaks, and coincident with a nuclear X-ray source within uncertainty \cite{JB05}. This structure can be explained if the continuum source, denoted by the ``core'' in Fig. 1f, is surrounded by an inclined molecular gas torus (diameter \(\sim61\) pc if the galaxy is at distance 10.8 Mpc). The core is not at the center of the torus, but offset by \(\sim10\) pc; this structure is similar to the nucleus of the Seyfert galaxy NGC 1068 \cite{GB16}. The scenario of a rotating torus is supported by the velocity information inferred from the CO data cube. The molecular gas in the core exhibits a separate velocity component, possibly due to presence of a massive central object (supermassive black hole, nuclear cluster), warped disk, or gas inflow/outflow.

\subsection{High-resolution CO excitation image of the starburst region}

Using our new CO (3-2) data and the CO (1-0) data from cycle 1, we derived an intensity ratio of CO (3-2) to CO (1-0), defined as \(R_{31}\equiv I_{32}/I_{10}\), where \(I\) is the integrated intensity in brightness units (Fig. 2a). Both CO images were produced by combining the data acquired with the 12-m array, Atacama Compact Array, and Total Power array, thereby recovering the total flux. The CO (3-2) image was smoothed to the angular and velocity resolution of CO (1-0). Fig. 2a shows that \(R_{31}\) is high (\(\sim1\)) in the central 200 pc, reaching a peak of \(R_{31}\simeq1.0\) in the galactic center (Fig. 2b); this region coincides with the SF disk revealed with radio continuum observations \cite{Sal16}. The 500 pc ring on average exhibits a much lower ratio of \(R_{31}\sim0.6\), decreasing to \(R_{31}<0.4\) in the outermost parts. Higher CO gas excitation in SF regions can be explained as a consequence of higher gas density and kinetic temperature, typical for starburst environments. For instance, a similar trend of decreasing \(R_{31}\) with radius has also been found in the starburst galaxy M82 \cite{Sal13}.

From the CO excitation image (Fig. 2a), we find \(R_{31}\simeq0.8\) in the GMCs in the inner 400 pc (SF GMCs in Fig. 1c). Using a conversion factor of \(X_\mathrm{CO}=0.8\times10^{20}~\mathrm{cm}^{-2}~(\mathrm{K~km~s}^{-1})^{-1}\), derived by applying radiative transfer modeling on the central 1 kpc \cite{Sal14}, we obtain GMC masses of the order of \(\sim1\times10^7~M_\odot\). The ISM mass of the same order of magnitude is obtained from the 0.87 mm continuum, with an assumed dust temperature of \(T_\mathrm{d}=25\) K, and applying the calculation method in \citet{Sco14}. An order of magnitude higher \(T_\mathrm{d}\) would yield approximately an order of magnitude lower ISM mass, implying an even lower value of \(X_\mathrm{CO}\). In either case, the conversion factor in the starburst nucleus of NGC 1808 is at least a factor of two lower than the standard Galactic value \cite{Bol13}, and the GMC masses are in the range of \(10^6\)-\(10^7~M_\odot\). This is comparable to the masses of super star clusters in galaxy centers.

\subsection{Dense gas tracers HCN (1-0) and HCO\(^{+}\) (1-0)}

We also observed the dense gas tracers HCN (1-0) and HCO\(^{+}\) (1-0) at a resolution of \(1''\) (\(\sim50\) pc). Both lines were detected throughout the central 1 kpc, exhibiting an overall structure similar to CO (3-2). In Fig. 2c, we show the distribution of the line intensity ratio of HCN (1-0) to HCO\(^{+}\) (1-0), \(R_\mathrm{dg}\equiv I_\mathrm{HCN}/I_\mathrm{HCO^+}\). This ratio has been used as a diagnostic tool for the physical conditions of molecular gas in active galactic nuclei (AGN), where a high \(R_\mathrm{dg}\) has been explained in terms of density and temperature effects, X-ray dominated chemistry, shocks, and infrared pumping \cite{Koh01,Ima06,Mei06,Har10,Aal12,Izu13,Taf10}. The ratio was imaged in NGC 1808 before, although at low resolution \cite{Gre16}. Averaged over image pixels, the ratio is \(R_\mathrm{dg}=1.45\pm0.08\) in the CND region and \(R_\mathrm{dg}=0.90\pm0.03\) in the 500 pc ring. The variation of the ratio is shown in Fig. 2e as an azimuthally averaged radial profile derived at a resolution of 1.5 arcseconds. Throughout the 500 pc ring and outer regions the ratio is close to unity. However, Fig. 2c shows that \(R_\mathrm{dg}\) is not constant in the ring, where values between \(0.5\) and \(1.3\) can be found. The ratio is not constant in the CND either; Fig. 2c,d shows that although the value in the galactic center is relatively high (\(1.5\)), similar to what has been observed recently in the Seyfert galaxies NGC 1068 and NGC 1097 \cite{GB14,Mar15}, the ratio exhibits a secondary peak in a ring-like structure with a radius of \(r\sim300\) pc. There are important implications from this result: (1) We found no correlation between \(R_\mathrm{dg}\) and X-ray luminosity. While X-ray regulated chemistry has been proposed to explain the elevated \(R_\mathrm{dg}\) in Seyfert nuclei, the absence of correlation in NGC 1808, which harbors a hard X-ray source (AGN candidate), implies that \(R_\mathrm{dg}\) is dominated by another process. (2) \(R_\mathrm{dg}\) can be enhanced in regions of higher density and temperature, including shocks. The gradient of \(R_\mathrm{dg}\) in Fig. 2d across the major axis of the CND (position angle \(155^\circ\)) indicates that \(R_\mathrm{dg}\) is elevated in the outer regions of the CND, where molecular gas is exposed to the feedback from the circumnuclear SF, although optical depth effects may be important in this region. Applying the diagnostic method of \citet{Koh01} and \citet{Ima06}, the nucleus of NGC 1808 lies in the boundary region between ``pure AGN'' and ``starburst'' classes, possibly a composite.

\subsection{Dense gas tracers CS (2-1) and SiO (2-1)}

In Fig. 2f, we show the distributions of CS (2-1) and SiO (2-1), observed at 98.0 and 86.8 GHz, respectively. While CS (2-1) is a reliable dense gas tracer, SiO (2-1) is known as a tracer of fast shocks, probing the regions of dense (\(n_\mathrm{H}\sim10^5\) cm\(^{-3}\)) molecular gas where the sputtering of dust grains occurs \cite{MP92,Sch97}.

Note that CS (2-1) in the CND exhibits the same double peak structure as CO (3-2), apparently coexisting in the molecular torus. On the other hand, SiO (2-1), detected for the first time in NGC 1808, is more centrally distributed, suggesting the presence of shocked molecular gas in the inner regions of the CND in the vicinity of the continuum core. The shocks may have been generated by intense SF activity in the nucleus, resulting in supernova explosions, as well as inflows and outflows of gas \cite{Sal16,Bus16}. Although SiO (2-1) emission could also be a product of an X-ray dominated region (XDR), as discussed for the case of NGC 1068 \cite{GB10}, we prefer the shock scenario because the galaxy center in NGC 1808 seems to be dominated by SF activity with evidence of noncircular motions and large velocity dispersion.

Further investigations that include more dense gas tracers and their isotopologues, as well as evaluation of optical depths and column densities will clarify the physical conditions and abundance variations of different molecular species in GMCs. NGC 1808 will remain an interesting target galaxy for further studies with ALMA.

\section{Concluding Remarks}

These first high-resolution, multi-line observations of NGC 1808 have given us an insight into the SF process and its feedback in a nearby starburst galaxy, demonstrating the revolutionary capabilities of ALMA. While the achieved high resolution yielded a fascinating view on the nuclear spiral pattern, distribution of individual GMCs, and the resolved intensity ratio of different molecular lines, the high sensitivity allowed detections of weak lines such as SiO (2-1), essential to understand the physical conditions of molecular gas. A more comprehensive presentation of the data will be given in a forthcoming paper (Salak et al. in prep.), while future ALMA observations at higher resolution will be able to resolve the core and reveal its structure and kinematics.

\section*{Acknowledgments}

This paper makes use of the following ALMA data: ADS/JAO.ALMA\#2013.1.00911.S. ALMA is a partnership of ESO (representing its member states), NSF (USA), and NINS (Japan), together with NRC (Canada) and NSC and ASIAA (Taiwan), in cooperation with the Republic of Chile. The Joint ALMA Observatory is operated by ESO, AUI/NRAO, and NAOJ. The National Radio Astronomy Observatory is a facility of the National Science Foundation operated under cooperative agreement by Associated Universities, Inc. D.S. was supported by the ALMA Japan Research Grant of NAOJ Chile Observatory, NAOJ-ALMA-59 and NAOJ-ALMA-98.

\begin{figure*}
\vskip -0.8cm
\centering
$\begin{array}{cc}
\includegraphics[angle=0,height=8.cm]{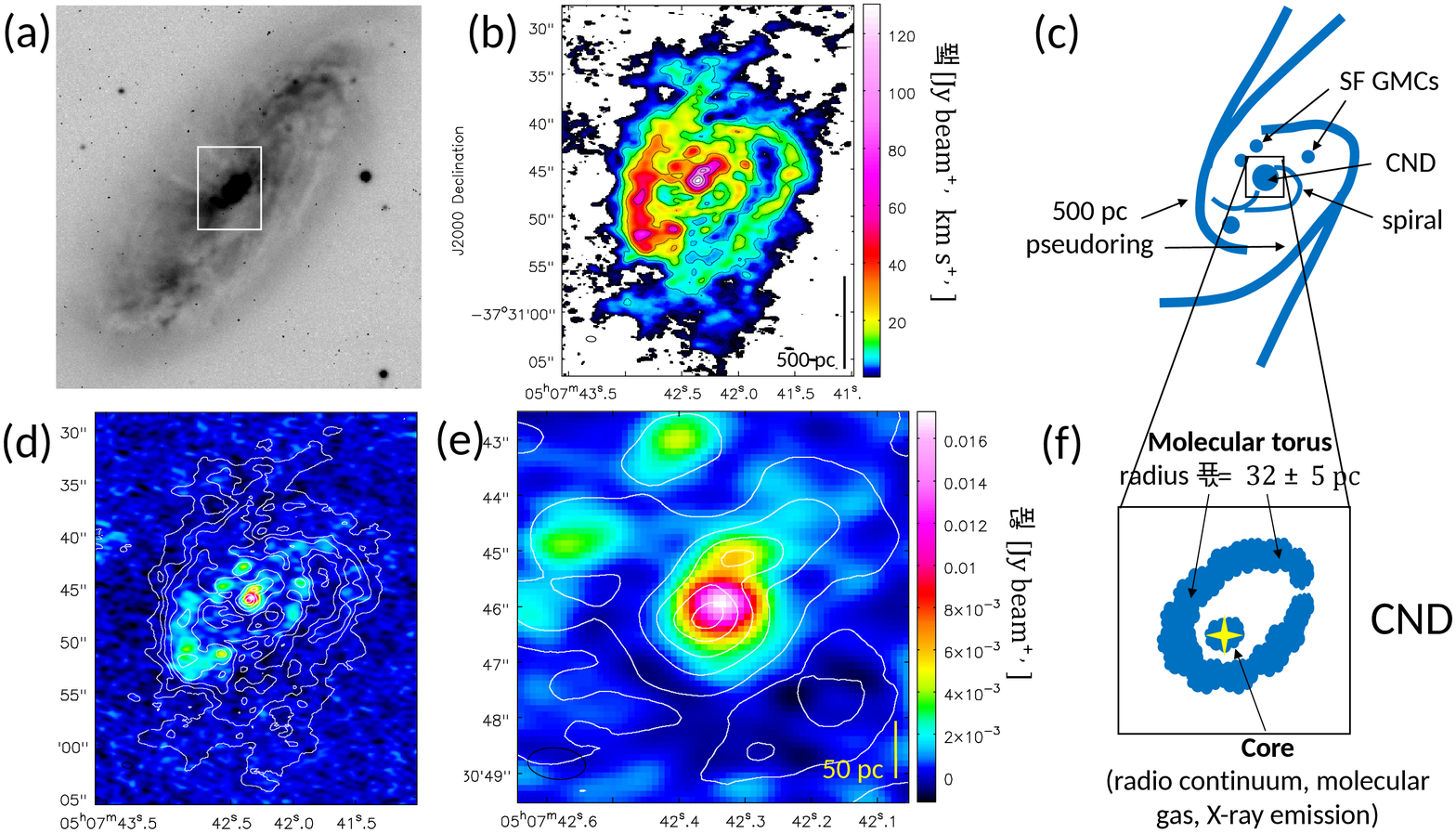} 
\end{array}$
\caption{\textbf{(a)} An optical image (\(B\) band) of NGC 1808 (NED). \textbf{(b)}  CO (3-2) contours (0.01, 0.05, 0.1, 0.2, 0.4, 0.6, 0.8, 0.95) times the maximum within the rectangle region in panel (a). \textbf{(c)} Illustration of the main structures of the CO (3-2) distribution in panel (b). \textbf{(d)} 0.87-mm continuum (color) with CO (3-2) contours. \textbf{(e)} Enlargement of panel (d). \textbf{(f)} Illustration of the molecular gas distribution in the circumnuclear disk (central 100 pc). The ``core'' exhibits radio, X-ray, and CO emission, marking the location of the AGN, whereas the double peak seen in panel (e) indicates the presence of a gaseous torus. The 0.87 mm continuum peak coincides with a hard X-ray source within the uncertainty.
}
\label{fig1}
\end{figure*}

\begin{figure*}
\vskip -0.8cm
\centering
$\begin{array}{cc}
\includegraphics[angle=0,height=8.cm]{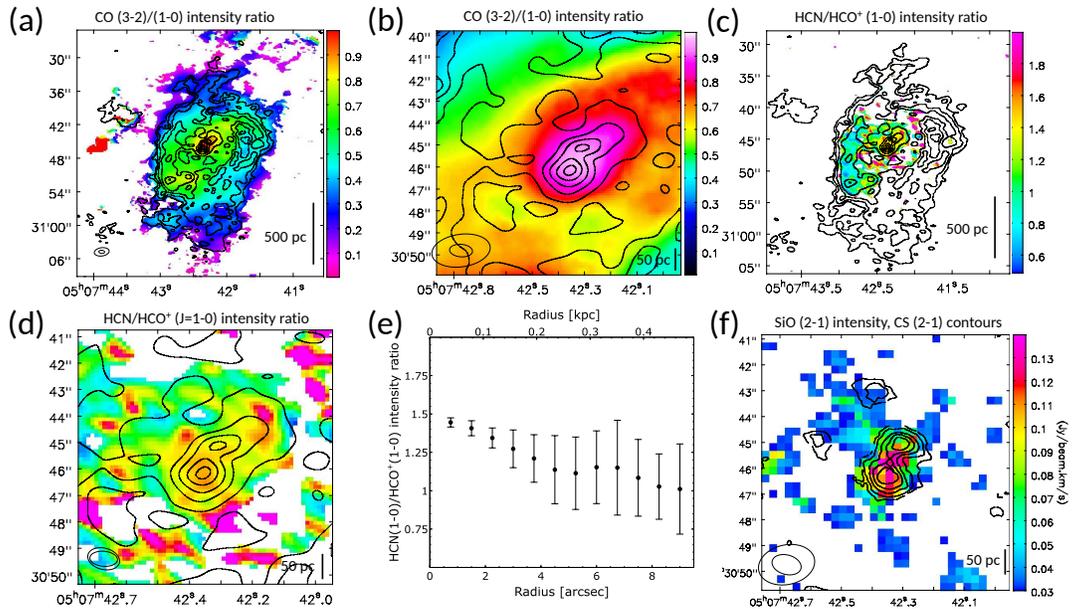} 
\end{array}$
\caption{\textbf{(a)} The intensity ratio of the CO (3-2) and CO (1-0) lines (each line intensity in brightness units; clipped at \(5~\sigma\)). The ratio image was derived after adjusting the angular resolution (\(2''\) or \(\sim100\) pc); CO (3-2) contours at (0.025, 0.05, 0.1, 0.2, 0.4, 0.6, 0.8, 0.95) times 133.4 Jy beam\(^{-1}\) km s\(^{-1}\). \textbf{(b)} Enlargement of figure (a). \textbf{(c)} The intensity ratio of the HCN (1-0) and HCO\(^{+}\) (1-0) lines at \(1''\) resolution with CO (3-2) contours. \textbf{(d)} Enlargement of (c). \textbf{(e)} The azimuthally averaged radial profile of the HCN (1-0) to HCO\(^{+}\) (1-0) ratio at a resolution of 1.5 arcsecs. \textbf{(f)} SiO (2-1) integrated intensity and CS (2-1) contours (0.1, 0.2, 0.4, 0.6, 0.8 times 0.744 Jy beam\(^{-1}\) km s\(^{-1}\)). The black cross marks the location of the radio continuum core.
}
\label{fig2}
\end{figure*}

\end{document}